\documentclass[12pt]{iopart}
\newcommand{\PeM}{{\rm PM}}
\newcommand{\MM}{{\rm MM}}
\newcommand\halfzeta{\mbox{$\frac{1}{2}$}\zeta}
\usepackage{graphicx}

\begin{document}

\title[Adsorption models of oligonucleotide microarrays]{Adsorption models of hybridisation and 
post-hybridisation behaviour on oligonucleotide microarrays}  

\author{Conrad J.\ Burden$^{1,2}$, Yvonne Pittelkow$^1$ and Susan R.\ Wilson$^1$}
\address{$^1$Centre for Bioinformation Science,
      Mathematical Sciences Institute\\
      Australian National University,
      Canberra, ACT 0200, Australia}
\address{$^2$John Curtin School of Medical Research\\
      Australian National University,
      Canberra, ACT 0200, Australia}
\ead{Conrad.Burden@anu.edu.au}

\begin{abstract}
Analysis of data from an Affymetrix Latin Square spike-in experiment indicates 
that measured fluorescence 
intensities of features on an oligonucleotide microarray are related to spike-in RNA target 
concentrations via a hyperbolic response function, generally identified as a 
Langmuir adsorption isotherm.  Furthermore the asymptotic signal at high spike-in concentrations 
is almost invariably lower for a mismatch feature than for its partner perfect match feature.  
We survey a number of theoretical adsorption models of hybridization at the microarray surface 
and find that in general they are unable to explain the differing saturation responses of perfect and
mismatch features.  On the other hand, we find that a simple and consistent explanation can be found in a model in which equilibrium hybridization followed by partial 
dissociation of duplexes during the post-hybridization washing phase.  
\end{abstract}


\maketitle

\section{Introduction}
\label{sec:Intro}

Oligonucleotide microarrays are designed to enable the evaluation of simultaneous expression 
of large numbers of genes in prepared messenger RNA samples.  Details of the technology and the 
design and manufacture of Affymetrix GeneChip arrays, the focus of this paper, can be found in 
the review of Nguyen et al.\cite{Nguyen02} or at the Affymetrix website 
{\tt http://www.affymetrix.com/technology/index.affx}.  The purpose of this paper is to examine 
physical models of hybridization of RNA at the microarray surface in the light of differing responses 
of perfect match and mismatch probes.  

In the manufacture of Affymetrix arrays, single strand DNA probes, 25 bases in length are 
synthesized base by base onto a quartz substrate using a photolithographic process.  They are 
attached to the substrate via short covalently bonded linker molecules roughly 10 nanometres apart.  
A microarray 
chip surface is divided into some hundreds of thousands of regions called features, commonly 11 to 20 
microns square, and with the single strand DNA probes within each feature being synthesized to a 
specific nucleotide sequence.  

A key step in the laboratory process of gene detection with microarrays is the 
hybridization of cRNA target molecules fractionated to lengths of typically 50 to 200 bases 
onto the single strand DNA probes.  The density of hybridized probe-target duplexes in each feature 
is detected via intensity measurements of fluorescent dye attached to the target cRNA molecules.  
Each gene or EST is represented by a set of 11 to 20 (dependent on the chip type) 
pairs of features using sequences of length 25 selected for their predicted hybridization properties and 
specificity to the target gene.  The first element of the pair, termed the perfect match (PM), 
is designed to be an exact match to the target sequence, while the second element, the 
mismatch (MM), is identical except for the middle (13th) base being replaced by its complement.  

A number of studies have demonstrated the appropriateness 
of Langmuir adsorption theory for understanding probe-target hybridization at the surface of microarrays.  
Experimental work includes that of Nelson et al.\cite{Nelson01}, 
Peterson et al.\cite{Peterson01,Peterson02} and Dai et al.\cite{Dai02}.  Analyses which 
have sought to match Langmuir adsorption isotherms with data from an Affymetrix spike-in 
experiment include those of Held et al.\cite{Held03}, Hekstra et al.\cite{Hekstra03}, 
Lemon et al.\cite{Lemon03}, Burden et al.\cite{Burden04} and Binder et al.\cite{Binder04}.  

The ultimate aim of such work is to establish a functional relationship between measured 
fluorescence intensities and underlying target 
concentration parameterized by known physical properties such as probe base sequences.  
If such a relationship could be established, it would offer the possibility of an absolute 
measure of RNA target concentration, as opposed to an arbitrarily defined `expression measure'.  
Fundamental to establishing this relationship is a model which accurately describes 
the physics of the various steps involved in producing a set of intensity measurements from a given mRNA 
target concentration.  The two steps we focus on in this paper are hybridization at the microarray 
surface and the subsequent washing step, designed to removed unbound target molecules.  

A little recognized shortcoming of existing hybridization models based on Langmuir adsorption theory is their 
inability to explain the differing responses of PM and MM fluorescence intensity signals at saturation 
concentrations of RNA.  That the asymptotic response of a MM feature at high PM-specific spike-in 
concentration should be less than that of the neighbouring PM feature is hardly news to an experimental 
biologist, and yet this observation is surprisingly difficult to reconcile with Langmuir adsorption theory 
(see Section~\ref{sec:Inconsistency}).  This problem was discussed in the early experimental work of 
Forman et al.\cite{Forman98}, 
who serendipitously recognized the `unexpected benefit' of the phenomenon of differential response between 
PM and MM, but failed to find a satisfactory physical explanation.  It is stated in the manufacturer's web page that  
`The reason for including a MM probe is to provide a value that 
comprises most of the background cross hybridization and stray signal affecting the PM 
probe.  It also contains a portion of the true target signal.'\cite{Affy02}   Consequently, 
many researchers have come to view 
the MM signal as primarily an attempt to measure non-specific hybridization and other background signal, 
though in practice there are problems with using the MM signals for this purpose\cite{Irizarry03}.  
Since the MM signals are more than a measure of non-specific hybridization, we will concentrate in 
this paper on the view that MM features are primarily less responsive versions of the PM features, and 
seek to understand their differing responses at saturation.  
The difference between PM and MM probe signals can then be exploited as the result of a single, well 
controlled change in one of the many parameters influencing the complicated process of hybridization.  
From this perspective one can obtain powerful insights into the physics and chemistry of 
hybridization at the microarray surface.  

In Section~\ref{sec:Langmuir} we review the Langmuir or hyperbolic isotherm and its relationship 
to a well known Affymetrix spike-in data set.   Section~\ref{sec:PhysMod} concentrates on an 
extension of the adsorption based hybridisation models of 
Hekstra et al.\cite{Hekstra03} and Halperin et al.\cite{Halperin04} which include the effects of 
non-specific hybridization, and which we show to be essentially equivalent to each other.  This 
model is consistent with a hyperbolic response function, as observed in data from spike-in experiments.  
However, as we point out in Section~\ref{sec:Inconsistency}, it is unable to explain the observed difference between PM and MM signals at saturation concentrations.  Section~\ref{sec:OtherEffects} 
is a survey of a number of possible improvements to our starting model of hybridisation at the microarray 
surface, which seek to overcome this shortcoming.  Many of these ideas have been 
canvassed in the literature, though in general they have not been rigorously examined in the light of 
the Hekstra/Halperin model.  In general, we find no convincing way of explaining the PM/MM 
difference at saturation by reference only to the hybridisation step.  In Section~\ref{sec:Washing} 
we consider the post-hybdridisation washing step, and find this to be the most promising 
explanation for the PM/MM difference.  
In Section~\ref{sec:Summary} we summarize our findings and draw conclusions.  Many of the technical calculations are relegated to appendices.  

\section{The Langmuir isotherm model}
\label{sec:Langmuir}

Langmuir adsorption theory is based on an assumption that there are two competing processes 
driving hybridization: adsorption, i.e.\ the binding of target molecules to immobilized probes 
to form duplexes, and desorption, i.e.\ the reverse process of duplexes dissociating into separate 
probe and target molecules
\begin{equation}
{\rm Probe} + {\rm Target} \rightleftharpoons {\rm Duplex}.   \label{basicreaction}
\end{equation} 
Herein we shall always use the word `probe' to indicate single strand DNA immobilised on the 
microarray, `target' to indicate RNA in solution and `duplex' to indicate a bound probe-target 
pair.  Both the forward and reverse processes are determined by chemical rate constants 
which depend on a number of factors including activation energies and temperature.  
Adsorption models of microarrays often lead to a hyperbolic response function, 
or equilibrium Langmuir isotherm, relating RNA target concentration $x$ to a measured equilibrium 
fluorescence intensity $y$, namely
\begin{equation}
y(x) = y_0 + b \frac{x}{x + K}.       \label{hypres}
\end{equation}
The isotherm is defined by three parameters: $y_0$ is the measured background intensity at zero 
target concentration, $b$ is the saturation intensity above background at infinite target 
concentration, and $K$ is the target concentration required to reach half saturation.  The 
physical origins of these parameters will be discussed in detail below.  

In a previous paper we have carried out an extensive statistical analysis \cite{Burden04}  
of fits of the hyperbolic and other response functions to the PM probes in the publicly available 
data from the Affymetrix Human HG-U95A Latin Square spike-in experiment 
({\tt http://www.affymetrix.com/support/technical/sample\_data/datasets.affx}).  
In this experiment genes (or, more precisely, RNA transcripts) were spiked in at cyclic
permutations of the set of known concentrations, together with 
a background of cRNA extracted from human pancreas.  The data consists of fluorescence intensity 
values from a set of 14 probesets corresponding to 14 separate genes, each 
containing 16 probe pairs.  For each probeset a set of fluorescence intensity values was 
obtained for the 14 spiked-in concentrations (0, 0.25, 0.5, 1, 2, 4, ..., 1024) pM.  The experiment 
was replicated three times using microarray chips from different wafers.  In common with previous 
analyses of this data set, our study concentrated on data from 12 of the 14 genes, omitting data from 
two defective genes.  

\begin{figure*}
\centering
\includegraphics{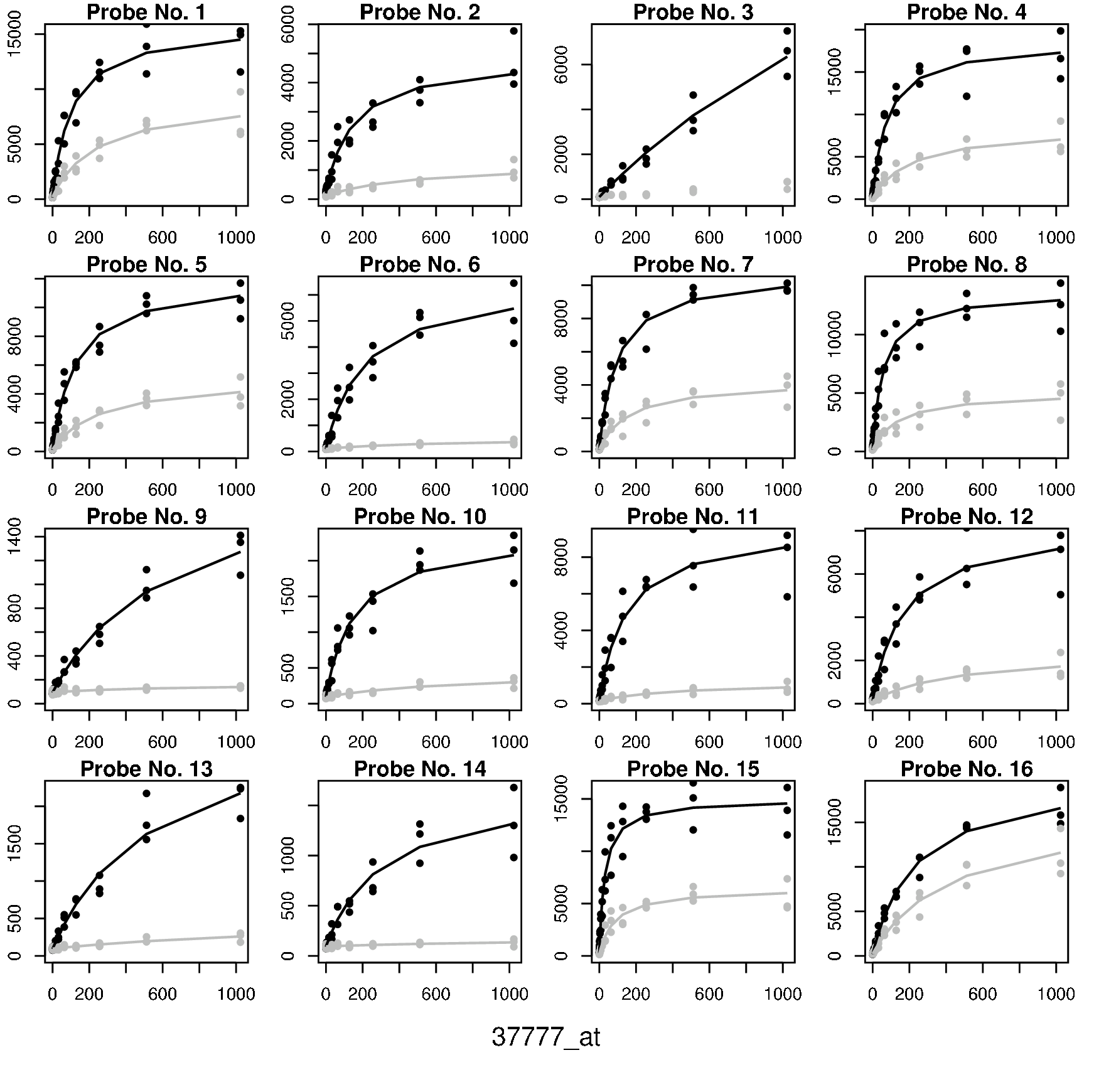}
\caption{Fits of Eq.~(\ref{hypres}) to fluorescence intensity data for the 16 PM (black) 
  and 16 MM (grey) features 
  of the gene 37777\_at probeset of the Affymetrix spike-in experiment.  Concentrations 
  (horizontal axes) are in picomolar and fluorescence intensities (vertical axes) are in the 
  arbitrary units used in Affymetrix .cel files.  The fit to MM probe No. 3 gave unphysical 
  negative values to the parameters $K$ and $b$ and is not shown.}
\label{fig:isotherms}
\end{figure*}

Fits of a number of functions to the fluorescence intensities were compared using 
a rigorous statistical analysis.  The optimum model of those considered for this data set 
is summarised as follows: 
\begin{enumerate}
\item Measured fluorescence values can be approximated by a Gamma distribution with 
mean given by Eq.~(\ref{hypres}) and constant coefficient of variation, here $\approx 0.17$.
\item The equilibrium isotherm Eq.~(\ref{hypres}) tracks fold changes from both 
PM and MM probes over the range of spiked-in concentrations from $<1$pM to $>1000$pM. 
\item All three parameters $y_0$, $b$ and $K$ are probe sequence dependent (in contrast 
with the findings of ref.~\cite{Held03}).
\item MM features almost invariably saturate at a lower asymptotic intensity $y_0 + b$ than their PM 
counterparts.
\end{enumerate}
Plots of fits of the hyperbolic response function to intensity data from the 16 PM and MM features corresponding to a typical one of the 12 genes considered is reproduced from ref.~\cite{Burden04}
in Fig.~\ref{fig:isotherms}.  A measure of the closeness of the fit is the unscaled deviation, defined 
by Eq.~(8) of ref.~\cite{Burden04}.  This quantity is the analogue for a generalized linear model 
of the mean square error in a standard linear regression.  For each of the 12 genes in question, 
the unscaled deviation 
per degree of freedom is much the same, the gene shown in Fig.~1.3 being somewhere near the
middle of the range.  Since the complete set of models considered in ref.~\cite{Burden04} 
was a set of nested models, we were able to 
use standard statistical tests based on accepted principles of balancing accuracy and parsimony to 
reject alternative functional forms for the fluorescence intensity response function, in favour of 
the hyperbolic form of Eq.~(\ref{hypres}).  The rejected response functions included 
a Sips isotherm\cite{Sips48} and a function modelling non-equilibrium adsorption 
(see Eq.~(\ref{tsoln})).  While details of the analysis were only reported for PM 
features in our earlier paper, we have subsequently also confirmed points (i) to (iii) for MM features 
(see Appendix A for comparison of hyperbolic and Sips isotherms).  Point (iv) was confirmed by 
fits of the hyperbolic response function by Hekstra et al. (see Fig.~2A of Ref.~\cite{Hekstra03} and 
accompanying text) and our own calculations, and is apparent from Fig.~\ref{fig:isotherms}. 

\section{Physical models leading to the hyperbolic isotherm}
\label{sec:PhysMod}

In what follows we define `specific' to mean PM specific.  All other hybridization will be 
referred to as `non-specific'.  Hekstra et al.\cite{Hekstra03} have modelled hybridization at the 
microarray surface in the combined presence of a specific cRNA target species and a single, non-specific target species using classical chemical adsorption 
kinetics.  The model gives a hyperbolic response function of the form Eq.~(\ref{hypres}) and 
predicts values for the parameters $y_0$, $b$ and $K$ in terms of chemical rate constants and 
physical properties of the microarray.  It is straightforward to extend their results to any number 
of non-specific species\cite{Binder04}.  

The hyperbolic isotherm is equivalently derivable from statistical mechanics by considering 
the Gibbs distribution at constant chemical potential\cite{Hill60}.  Halperin et al.\cite{Halperin04} 
have used this approach to study adsorption in microarray chips in the presence of non-specific 
hybridization.  In order to establish a notation for subsequent sections, we rederive here 
the hyperbolic isotherm using the Halperin approach.  We shall further augment the approach 
to include partial zippering of duplexes, that is, the idea is that a particular probe-target 
duplex can exist in a number of possible partially zipped-up  configurations 
$\alpha = 1, 2, \ldots$(see, for example, ref.~\cite{Deutsch04}).  

For a given feature on the microarray surface, whether PM or MM, 
let the concentration of target molecules specific to the PM feature of the matched pair be $x$, 
and the concentration of the non-specific species $i$ be $z_i$.   Further, 
let $\theta_\alpha$ be the fraction of a given feature covered by specific duplexes in partially zippered 
configuration $\alpha$, and likewise $\phi_{i\alpha}$ be the fraction covered by duplexes formed with 
non-specific target species $i$ in configuration $\alpha$.  The fraction covered with unmatched single strand probes is  therefore $1 - \theta - \sum_i \phi_i$, where the total fraction of sites 
holding respectively specific and non-specific duplexes of the $i$th species is 
$\theta = \sum_\alpha \theta_\alpha$ and 
$\phi_i = \sum_\alpha \phi_{i \alpha}$. The free energy per mole 
of probe sites at the microarray surface is 
\begin{eqnarray}
\gamma & = & RT \left[ \sum_\alpha \theta_\alpha \ln\theta_\alpha 
                     + \sum_{i,\alpha} \phi_{i \alpha} \ln\phi_{i \alpha} + 
   \left(1 - \theta - \sum_i \phi_i\right)
          \ln\left(1 - \theta - \sum_i\phi_i\right) \right]  \nonumber \\
  & & + \; \sum_\alpha \theta_\alpha \mu_{{\rm pt}\alpha}^0 + 
                           \sum_{i,\alpha} \phi_{i\alpha} \mu_{{\rm pt}i\alpha}^0 
           + \left(1 - \theta - \sum_i \phi_i\right) \mu_{\rm p}^0,    \label{freeenergy}
\end{eqnarray}
where $\mu_{{\rm pt}\alpha}^0$, $\mu_{{\rm pt}i\alpha}^0$ and $\mu_{\rm p}^0$ are
respectively reference state chemical potentials per mole of specific and non-specific 
probe-target duplexes in configuration $\alpha$, and of unmatched probes
\footnote{Halperin et al.\cite{Halperin04}  also include a term for the charge density dependent electrostatic free energy, which we discuss briefly in Section~\ref{sec:ESBlocking}.} .  $R$ is the gas 
constant and $T$ the absolute temperature.  
The exchange chemical potentials of the various species of probe-target duplexes are 
\begin{eqnarray}
\frac{\partial \gamma}{\partial \theta_\alpha} & = & RT \left[\ln\theta_\alpha - 
       \ln\left(1 - \theta - \sum_i\phi\right)\right] 
                      + \mu_{{\rm pt}\alpha}^0 - \mu_{\rm p}^0,   \nonumber \\
\frac{\partial \gamma}{\partial \phi_{i\alpha}} & = 
          & RT \left[\ln\phi_{i\alpha} - \ln\left(1 - \theta - \sum_i\phi\right)\right] 
                      + \mu_{{\rm pt}i\alpha}^0 - \mu_{\rm p}^0.   \nonumber
\end{eqnarray}
At equilibrium these exchange chemical potentials 
balance the chemical potentials of the corresponding target molecule species in 
solution.  Assuming the bulk concentrations of target molecules are not appreciably affected by 
hybridization, these are given in terms of reference values $\mu_{\rm t}^0$ and $\mu_{{\rm t}i}^0$ 
at reference concentrations $x_0$ and $z_{0i}$ of specific and non-specific target molecules by 
\begin{eqnarray}
\mu_{\rm t} & = & \mu_{\rm t}^0 + RT \ln\frac{x}{x_0},  \label{targetpot} \\
\mu_{{\rm t}i} & = & \mu_{{\rm t}i}^0 + RT \ln\frac{z_i}{z_{0i}}.    \nonumber
\end{eqnarray}
Matching exchange chemical potentials with target chemical potentials gives 
\begin{eqnarray}
RT \ln\frac{x}{x_0} & = & RT \left[\ln\theta_\alpha - 
              \ln(1 - \theta - \sum_i\phi) \right] + \Delta G_\alpha, \nonumber \\
RT \ln\frac{z_i}{z_{0i}} & = & RT \left[\ln\phi_{i\alpha} - 
               \ln(1 - \theta - \sum_i\phi) \right] + \Delta G_{i\alpha},           \nonumber
\end{eqnarray}
where we have defined the duplex binding free energies 
\begin{equation}
\Delta G_\alpha = \mu_{{\rm pt}\alpha}^0 - \mu_{\rm p}^0 - \mu_{\rm t}^0, \hspace{5 mm}
\Delta G_{i\alpha} = \mu_{{\rm pt}i\alpha}^0 - \mu_{\rm p}^0 - \mu_{{\rm t}i}^0.    \label{Galphadef}
\end{equation}

Solving for the duplex coverage fractions $\theta_\alpha$ and $\phi_{i\alpha}$, and summing 
over configuratons $\alpha$, we obtain the isotherms 
\begin{eqnarray}
\theta & = & \frac{x/K_{\rm S}}{1 + x/K_{\rm S} + \sum_i z_i/K_i} \label{theta} \\
\phi_i & = & \frac{z_i/K_i}    {1 + x/K_{\rm S} + \sum_j z_j/K_j},   \label{phi}
\end{eqnarray}
 where  $K_{\rm S}^{-1}$ and $K_i^{-1}$ are effective equilibrium 
constants for specific and nonspecific hybridisations given by 
\begin{equation}
K_{\rm S}^{-1} = x_0^{-1} \sum_\alpha e^{-\Delta G_\alpha/RT}, 
\hspace{5 mm} K_i^{-1} = z_{0i}^{-1} \sum_\alpha e^{-\Delta G_{i\alpha}/RT}.
                                                     \label{effectiveK} 
\end{equation}
Introducing proportionality constants $b_{\rm S}$ and $b_i$ for the specific and 
non-specific hybridizations and a physical optical background $a$, the measured fluorescence 
intensity is given by  
\begin{eqnarray}
y(x) = a + b_{\rm S} \theta + \sum_i b_i\phi_i \label{bSdef} \\ 
     = y_0 + b \frac{x}{x + K},           \label{hekmodel}
\end{eqnarray}
where
\begin{equation}
y_0 = a + A, \hspace{5 mm}   
  b = b_{\rm S} - A, \hspace{5 mm}
  K = K_{\rm S}B, \label{y0bKcorrection}
\end{equation}
and 
\begin{equation}
A = \frac{1}{B}\sum_i \frac{b_i z_i}{K_i}, \hspace{5 mm}
B = 1 + \sum_i \frac{z_i}{K_i}. \label{ABdef}
\end{equation}
The presence of non-specific hybridization does not spoil the hyperbolic form 
of the Langmuir isotherm Eq.~(\ref{hypres}), but does influence the parameters $y_0$, $b$ and $K$.  
The purpose of Eqs.~(\ref{effectiveK}), (\ref{y0bKcorrection}) and (\ref{ABdef}) is to relate the estimated 
isotherm parameters to the underlying physical parameters: $a$ (the physical background value in the 
absence of any hybridization), $b_{\rm S}$ and $b_i$ (proportionality constants relating the 
incremental change in measured intensity to an incremental change in duplex fraction for the specific 
and non-specific hybridizations respectively), 
duplex binding energies $\Delta G_\alpha$ and $\Delta G_{i\alpha}$, and a set of 
non-specific background target concentrations $z_i$.  The parameters $b_{\rm S}$ and $b_i$ are 
a measure of the amount of fluorescent light emitted per hybridized target molecule.  Fluorescent 
dye is bound only to the target molecules (in fact only to U and C bases), so $b_{\rm S}$ and 
$b_i$ can only be functions of specific and non-specific target sequences, and not probe sequences.  
Eqs.~(\ref{hekmodel}) to (\ref{ABdef}) are a generalisaton of Eq.~(2) of Hekstra et al.\cite{Hekstra03}

\section{Inconsistency of adsorption models with observed PM/MM saturation intensities}
\label{sec:Inconsistency}

The model given by 
Eqs.~(\ref{effectiveK}) to (\ref{ABdef}) inescapably leads to a conclusion that 
the PM and MM intensity measurements for a given probe pair must saturate at the same asymptotic 
intensity value, contradicting the observed fits to experimental data.  
This point has been inferred previously in regard to adsorption models\cite{Forman98}, 
but does not appear to be generally 
appreciated in the literature, with the exception of work by Peterson et al.\cite{Peterson02}.   

Consider two neighbouring features on a microarray, one PM and one MM, their probe sequences 
differing only by the middle base.  Recall that, in this paper, we define the 
word `specific' to mean those target cRNA which are exact complements to the PM sequence, even 
when dealing with the MM feature.  For our purposes, this definition will prove useful given 
that, for most probe pairs, the dominant part of the MM signal at high spike-in concentrations 
in the Affymetrix experiment appears to come from hybridization of spiked-in target RNAs complementary to the PM sequence.  Parameters relating to the PM and MM features will be indicated 
by superscripts PM and MM respectively.  

Although the sums occurring in Eq.~(\ref{ABdef}) will be over the same set of 
non-specific targets for PM as for MM, one can expect $A^\MM \ne A^\PeM$ since in general 
$K_i^\MM \ne K_i^\PeM$.  
Considering the asymptotic intensities at high concentration, however, Eqs.~(\ref{hekmodel})
and (\ref{y0bKcorrection}) 
imply that, under the Hekstra model, the non-specific hybridization effects cancel out: 
\begin{equation}
\begin{array}{rcccl}
y^\MM(\infty) & = & y_0^\MM + b^\MM & = & a + b_{\rm S},\\
y^\PeM(\infty) & = & y_0^\PeM + b^\PeM & = & a + b_{\rm S}. 
\end{array}                                          \label{naiveasymp}
\end{equation}
An essential step in this argument is the claim that the parameters $a$ and $b_{\rm S}$ do 
not differ between intensity measurements from a neighbouring PM/MM pair of features.  
For the physical background $a$ this is clearly a reasonable 
assumption: physical properties of the chip in the absence of any hybridization, such as reflectance, 
are unlikely to vary significantly over a distance of a few microns.  For the parameter $b_{\rm S}$
the argument is more subtle.  From Eq.~(\ref{bSdef}), $b_{\rm S}$ is, up to a multiplicative constant, 
the expected number of biotin labels per hybridized specific target molecule.  Importantly, $b_{\rm S}$ 
confers on Eq.~(\ref{bSdef}) no information about probe-target binding affinities, this information 
being contained in the coverage fraction $\theta$.  By our current definition of `specific', target 
molecules contributing to the specific part of the signals of a given PM/MM pair of features are 
drawn from the same subset of molecules in the RNA solution, namely those containing a 
contiguous PM-specific subsequence of 25 bases.  
Hence $b_{\rm S}$ is the same for both members of a 
neighbouring PM/MM pair.  The Hekstra or Halperin model formulated above then necessarily 
entails that $y_0^\MM + b^\MM = y_0^\PeM + b^\PeM$, 
in obvious contradiction with the values of $y_0$, and $b$ obtained by fitting the spike-in data.  

The source of the problem is that any model leading to 
the coverage fraction given by Eq.~(\ref{theta}) entails that at sufficiently 
high specific target concentration, all probes form duplexes: as $x\rightarrow \infty$, 
$\theta \rightarrow 1$.  That is, all probes in the feature are predicted to 
form duplexes if saturated with enough specific target, even in the case of the MM 
feature.  
A subtle point to note is that this is true irrespective of the bulk solution melting temperature of duplexes, which is defined as the temperature at which half the total number of single strand targets are free and half are bound as duplexes in bulk solution.  This temperature can be calculated\cite{Bhanot03} in terms of enthalpy and entropy by balancing forward and backward reaction rates under the constraints of stoicheiometry, namely: $2 [{\rm T}] + [{\rm T.T}] = \mbox{constant}$, where $[{\rm T}]$ and $[{\rm T.T}]$ are bulk concentrations of single strand and duplex targets respectively
\footnote{In Sec.~\ref{sec:CompBulk} we argue that, for the Affymetrix spike-in data set $[{\rm T}]\approx x$, the spike-in concentration.}. 
However, this stoicheiometric constraint does not apply for the adsorption reaction at the microarray surface: because the solution target volume is effectively infinite, the target concentration is unchanged as hybridisation proceeds and the coverage fraction theta increases towards its finite equilibrium value $\theta \le 1$ .  Even above the bulk solution melting temperature, the forward reaction can be forced by setting the target concentration sufficiently high.  The upshot is that, irrespective of temperature, Langmuir adsorption theory tells us that a feature will saturate at infinite target concentration.

\section{Other hybridisation effects}
\label{sec:OtherEffects}

Clearly the above model does not account for all possible effects during the complex process of hybridisaton.  In this section we consider a number of other possible hybridisation effects, 
some of which have been proposed in the literature as putative explanations for the differing 
measured PM/MM saturation intensities.  In general, we find none of these effects to be a strong 
candidate, and believe that the explanation of the PM/MM saturation difference is unlikely to lie with 
the hybridisation step.  

\subsection{Sips isotherm}
\label{sec:Sips}

The problem of differential PM/MM saturation was recognized in the context of a simple 
Langmuir model without non-specific hybridization by 
Peterson et al.\cite{Peterson02}, who explain their experimental data by 
invoking a Sips isotherm to explain a lower MM response curve at high target concentrations.  
The Sips isotherm\cite{Sips48} is an empirical response curve believed 
to correspond to an adsorption model in which chemical reaction rates are drawn from a 
pseudo-Gaussian distribution.  Peterson et al.'s experimental results are indeed a good fit to 
the Sips isotherm, however their experiment differs from the conditions of the hybridization of 
Affymetrix chips in one important aspect, namely the hybridization temperature.  The Peterson 
experiment was carried out at a hybridization temperature of 20$^\circ$C, while Affymetrix microarrays 
are hybridized at 45$^\circ$C.  Furthermore, 
Peterson et al.\ found that heating the hybridization buffer to 37$^\circ$C and then cooling back to 
20$^\circ$C almost completely removed any difference in equilibrium saturation intensities between 
PM and MM probes.  This appears to be the effect of a first order phase transition which sets in at a temperature well below the Affymetrix hybridisation temperature.  We comment on the problem of determining the phase structure in Sec.~\ref{sec:CompPP}.  

To determine whether the hyperbolic or Sips isotherm is more appropriate for the Affymetrix 
spike-in data we have carried out a statistical analysis comparing the fits of the MM data to both 
isotherms.  Our results, summarized in Appendix~A, show that for the Affymetrix spike-in data the 
extra parameters involved in invoking the Sips isotherm are not significant, and that a hyperbolic 
response function adequately describes the data.  We conclude that, at a hybridization 
temperature of 45$^\circ$C, the more appropriate empirical fit to the spike-in data is 
Eq.~(\ref{hypres}), with $y^\MM(\infty) < y^\PeM(\infty)$.  

\subsection{Non-equilibrium hybridization}
\label{sec:NonEq}

In an earlier paper~\cite{Burden04} we examined the possibility that hybridization had not 
reached equilibrium in the Affymetrix spike-in experiment.  We considered the simple 
non-equilibrium one-step model without non-specific hybridization, namely,  
\begin{equation}
\frac{d\theta}{dt} = k_{\rm f}x(1 - \theta) - k_{\rm b}\theta, 
                                                                  \label{onestep} 
\end{equation}
where $k_{\rm f}$ and $k_{\rm b}$ are forward and backward chemical reaction rates.  The solution
corresponding to the initial condition $\theta(x,0) = 0$ is 
\begin{equation}
\theta(x,t) = \frac{x}{x + K_{\rm S}} \left[1 - e^{-(x + K_{\rm S})k_{\rm f}t} \right],  \label{tsoln}
\end{equation}
where $K_{\rm S} = k_{\rm b}/k_{\rm f}$.  
A statistical analysis of the data showed that the extra degree of 
freedom distinguishing the non-equilibrium from the equilibrium solution (Eq.~\ref{hypres}) is 
not significant.  That is, our finding was that the equilibrium solution is the more appropriate model.  

However, textbook descriptions of duplex formation (see, for instance,  ref.~\cite{Cantor80}, pages 
1215 to 1219) imply that hybridization is more accurately described as a two step process: a slow rate 
determining step in which an initial two or three base pairs form, followed by a fast 
`zipping-up' step involving the remaining base pairs.  Measured forward reaction rates for 
duplex formation may typically be of the order of $10^6$ mol$^{-1}$sec$^{-1}$~\cite{Wang95}, 
potentially translating to timescales of several hours at picomolar concentrations.  In order to
establish more rigorously that the hybridization had reached equilibrium in the spike-in 
experiment, we have considered in Appendix~B a quasi-equilibrium hybridization model with 
two timescales.  Chemical reaction rates leading to the initiation configuration with two 
or three base pairs formed are taken to be slow, while other reaction rates are assumed to 
equilibrate on short timescales.  Again this model leads to non-equilibrium solutions taking the form of 
Eq.~(\ref{tsoln}), which differs from the hyperbolic form observed in the data.  This confirms that our 
previous statistical analysis is appropriate even when 
a two step hybridization process is taken into account.  We therefore believe that equilibrium 
thermodynamics to be the correct framework for studying hybridization for this dataset.  

\subsection{Electrostatic surface potential}
\label{sec:ESBlocking}

Halperin et al.\cite{Halperin04} include in the free energy Eq.~(\ref{freeenergy}) 
a term $\gamma_{\rm el}$ for the charge density dependent electrostatic free energy.  
The effect of this term is to change the effective equilibrium constants $K_{\rm S}$ and 
$K_i$ by a finite amount via the replacements 
$\Delta G_\alpha \rightarrow \Delta G_\alpha + \partial \gamma_{\rm el}/\partial \theta_\alpha$ 
and 
$\Delta G_{i\alpha} \rightarrow \Delta G_{i\alpha} + \partial \gamma_{\rm el}/\partial \phi_{i\alpha}$  in Eq.~(\ref{effectiveK}).  
This introduces a $\theta$ dependence to $K_{\rm S}$ and has the potential to change the shape of 
the isotherm from a hyperbolic form\cite{Pettitt02}.  However, it cannot be the explanation 
for differing PM/MM saturation intensities, as the adjusted form of Eq.~(\ref{theta}) still satisfies 
$\theta \rightarrow 1$ as $x\rightarrow \infty$.  

\subsection{Competitive hybridization with probe-probe pairs and probe self-interactions} 
\label{sec:CompPP}

Forman et al.\cite{Forman98} advanced the hypothesis that the observed divergence of 
saturation intensities between 
PM and MM features is caused by hybridization of neighbouring probe-probe pairs, rendering 
a certain fraction of each feature unavailable for target binding to form probe-target duplexes.  
Probe-probe interactions are possible if we assume probes of approximate length 8 nm and bound 
by flexible linker molecules to a glass substrate to have an average interprobe distance of the order of
10nm~\cite{Pirrung02}, especially given that some clustering of probes is to be expected.   
It has also been recognised\cite{Binder04} that self-interaction of probes via probe folding may 
render a fraction of probes unavailable for hybridization and affect adsorption isotherms.  

In Appendix~C we discuss how hybridizaton in the presence of probe-probe and probe self 
interactions may be modelled.  In agreement with 
ref.\cite{Binder04} we find that probe self-interactions have the effect of scaling the equilibrium 
constant $K_{\rm S}$ for the adsorption process.  However, this cannot be the 
explanation for differing PM/MM saturation intensities as it does not change the saturation 
asymptote.  The probe-probe interactions, on the other hand, are more complex, and we 
show in appendix~C that one is naturally led to the random lattice version of a two dimensional 
statistical mechanics model known as the monomer-dimer model.  No solution to this model 
exists even for the more tractable cases of regular lattices, though some numerical work has 
been done for the regular square lattice monomer-dimer model\cite{Baxter68}.  

\begin{figure}
\centering
\includegraphics[scale=0.8]{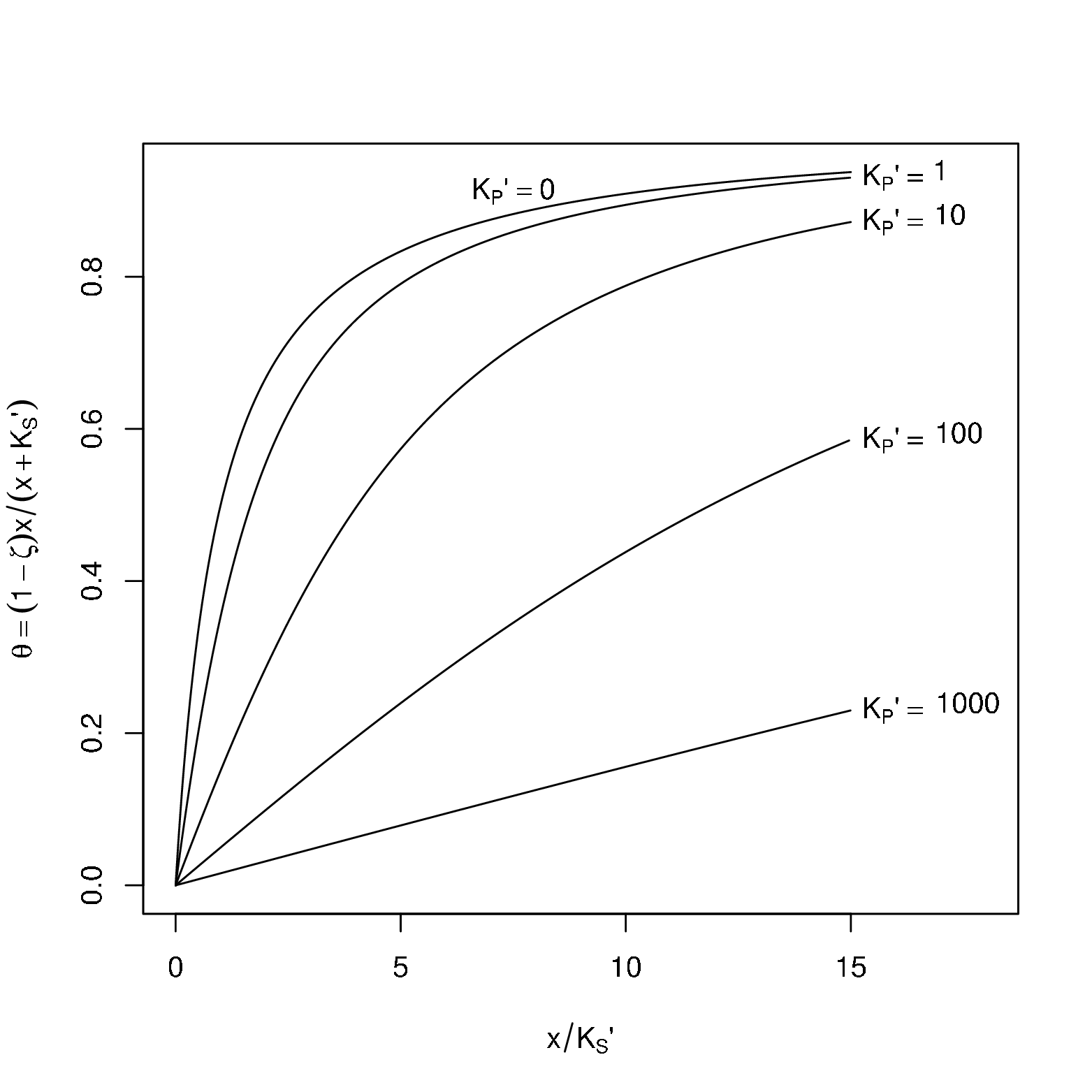}
\caption{Plots of the coverage fraction $\theta$ of probe-target duplexes against the dimensionless 
target concentration $x/K'_{\rm S}$ given by the solution Eqs.~(\ref{thetaPPprime}) and 
(\ref{thetazetaprime}) to the one-dimensional model described in Appendix~C, for various values 
of the effective probe-probe duplex equilibrium constant $K'_{\rm P} = K_{\rm P}/(1 + K_{\rm Q})^2$.  Probe duplex or probe self interaction free energies of 
$\Delta G_{\rm P}$ or  $\Delta G_{\rm Q} = 0$, $-1$ and $-2$ kcal~mol$^{-1}$ at $45^\circ$C 
correspond to $K_{\rm P}$ or $K_{\rm Q}$ values of 1, 4.9 and 23.7 respectively.}
\label{fig:onedim}
\end{figure}

In Appendix~C we tackle the unphysical but analytically tractable one dimensional 
model of competitive hybridization with probe-target and probe-probe duplexes.  We see that 
a probe-probe binding energy of 1 or 2 kcal mol$^{-1}$ is enough to make a noticeable difference to 
the adsorption isotherm in this approximation (see Fig.~\ref{fig:onedim}).  The one dimensional 
model saturates at 
100\% coverage of probe-target duplexes at high target concentration and so is unable to explain the 
divergence of PM and MM saturation intensities.  However it is well known that the behaviour of 
statistical mechanics models in one and two dimensions can be very different.  It is known, for instance, 
that a one dimensional model with local interactions cannot lead to a phase transition, whereas a number 
of two dimensional models are known to exhibit phase transitions at critical temperatures or 
densities\cite{Baxter82}.   

The evidence from numerical calculations of the monomer-dimer model on a regular square lattice 
is that it does not have a phase transition for non-zero monomer density\cite{Baxter68}, 
but we are unaware of any numerical 
simulations for the random lattice case more relevant to our problem.  
It is therefore still possible that the microarray surface configuration could undergo a 
phase transition from a disordered phase with low concentration of probe-probe duplexes to 
an ordered phase in which a high concentration of probe-probe duplexes line up along a particular 
direction.  This could explain the differing intensity measurement curves of MM features observed 
before and after quenching in the experiments of Peterson et al.\cite{Peterson02}.  Whether the Forman 
hypothesis can explain the observed difference in PM/MM saturation intensities, however, 
remains an open question, though any such function is unlikely to be consistent with the observed  hyperbolic response function.  

\subsection{Competitive bulk hybridization}
\label{sec:CompBulk}

By competitive bulk hybridization we mean the hybridization of specific target molecules T in solution 
either with (i) other specific target molecules T$'$ which might happen to be, at least in part, self 
complementary (${\rm T} + {\rm T} \rightleftharpoons {\rm T.T}$), (ii) non-specific target 
molecules which happen to have approximately complementary nucleotide sequences 
(${\rm T} + {\rm T'} \rightleftharpoons {\rm T.T'}$), or (iii) target self-interactions 
(${\rm T} \rightleftharpoons {\rm T_{folded}}$).  Halperin et al.\cite{Halperin04} have considered the 
effect on equilibrium isotherms of the first two types of bulk hybridization, and type (iii) can be dealt 
with in a similar way.  Assuming that 
probe-target hybridization has a negligible effect on bulk target concentrations, they argue 
that equilibrium isotherms can be obtained from isotherms such as Eq.~(\ref{theta}) by 
replacing the spike-in target concentration $x$ with the single strand concentration $[{\rm T}]$ 
obtained by applying the law of mass-action to the bulk hybridization reaction in solution.   

For all three types of hybridization, we argue here that competitive bulk hybridization cannot explain 
differential PM/MM saturation.  In each case, application of the law of mass action entails that 
$[{\rm T}] \rightarrow \infty$ as $x \rightarrow \infty$, so Eq.~(\ref{theta}) with $x$ replaced 
by [T] still implies 100\% saturation of features in the high spike-in concentration limit for 
both PM and MM features.  

Furthermore, we can rule out any significant effect on the isotherm from T.T hybridization 
for the probe sequences studied in the Affymetrix spike-in experiment by the following argument.  
The law of mass action implies that the behaviour of the 
single strand concentration goes from $[{\rm T}] \approx x$ 
at spike-in concentrations $x << K_{\rm bulk}^{-1}$ (where $K_{\rm bulk}$ is the equilibrium constant for the reaction 
${\rm T} + {\rm T'} \rightleftharpoons {\rm T.T'}$) to $[{\rm T}] \approx (x/2K_{\rm bulk})^{1/2}$ 
at spike-in concentrations $x >> K_{\rm bulk}^{-1}$.   A significant effect from T.T hybridization would 
therefore lead to a Sips isotherm with parameter $\gamma = \frac{1}{2}$ at high spike-in concentration, 
which, by the analysis of Appendix~A, is not observed over the range of concentrations in the Affymetrix 
spike-in experiment.  

\section{The washing step}
\label{sec:Washing}

The hybridization step is followed by a washing step designed to remove unbound target molecules before scanning the microarray.  During the washing step the target solution is flushed out of the cartridge containing the microarray and replaced by a washing buffer containing no RNA.  Thus 
the ambient concentration of target molecules 
is set to zero, switching off the forward adsorption reaction.  
We argue here that the washing step is responsible for 
the measured differences between PM/MM intensity measurements at saturation concentrations.  This 
idea has been proposed briefly by Zhang\cite{Zhang03a}, but requires further analysis.  

Let us assume that, immediately prior to washing, duplex coverage fractions on a given feature 
are given by the equilibrium 
model set out in Section~\ref{sec:PhysMod}.  That is, the fraction $\theta$ of sites 
on a feature occupied by specific probe-target duplexes and the fraction $\phi_i$ covered by 
non-specific duplexes of species $i$ are given by Eqs.~(\ref{theta}) and (\ref{phi}).  During the 
washing process some of the duplexes will be dissociated.  Suppose that the probability that a 
given probe-target duplex has survived up to a washing time $t_W$ is $s(t_W)$ for a specific 
duplex and $s_i(t_W)$ for a non-specific 
duplex of species $i$.  The survival functions $s$ and $s_i$ depend only on probe and target 
base sequences and not the ambient target concentrations $x$ and $z_i$ present during the 
prior hybridization step.  They satisfy $s(0) = 1$ and 
are monotonically decreasing.  The specific and non-specific duplex coverage fractions at time $t_W$ 
are then 
\begin{eqnarray}
\theta(x,t_W) & = & \frac{s(t_W) x/K_{\rm S}}{1 + x/K_{\rm S} + \sum_i z_i/K_i} \label{thetaW} \\
\phi_i(x,t_W) & = & \frac{s_i(t_W)z_i/K_i}    {1 + x/K_{\rm S} + \sum_j z_j/K_j}.   \label{phiW}
\end{eqnarray}

Repeating the assumption used in Section~\ref{sec:Langmuir} that the measured fluorescence intensity is a linear function 
of the duplex coverage fractions, that is 
$y(x,t_W) = a + b_{\rm S}\theta(x,t_W) + \sum_i b_i \phi(x,t_W)$, we find that at fixed $t_W$ 
the hyperbolic form required by points (ii) and (iii) in Section~\ref{sec:Langmuir}, namely  
\begin{equation}
y(x,t_W) = y_0(t_W) + b(t_W) \frac{x}{x + K},           \label{hekmodelW}
\end{equation}
is maintained, and that the `observed' parameters $y_0$, $b$ and $K$ are now given by 
\begin{equation}
y_0(t_W) = a + A(t_W), \hspace{1 mm}   
  b(t_W) = s(t_W) b_{\rm S} - A(t_W), \hspace{1 mm}
  K = K_{\rm S}B, \label{y0bKcorrectionW}
\end{equation}
where 
\begin{equation}
A(t_W) = \frac{1}{B}\sum_i \frac{s_i(t_W) b_i z_i}{K_i}, \hspace{5 mm}
B = 1 + \sum_i \frac{z_i}{K_i}. \label{ABdefW}
\end{equation}

Note that the parameter $K$ is unaffected by the length of the washing process, and depends only on duplex 
binding free energies via the hybridization step.  The asymptotic fluorescence intensity at high target 
concentration, 
\begin{equation}
y(\infty,t_W) = y_0(t_W) + b(t_W)  = a + s(t_W) b_{\rm S}, \label{yinfW}
\end{equation}
is depressed by the presence of the survival 
fraction $s(t_W)$.  

To model the survival function $s(t_W)$, one expects the rate of dissociation of specific probe-target 
duplexes to be the product of the fraction $\theta(t_W)$ of probes forming specific duplexes 
and a washing rate $\kappa$ which depends only on the probe and target nucleotide sequences.  
Assuming then that $\kappa$ is independent of $t_W$, the survival function is 
\begin{equation}
s(t_W) = e^{-\kappa t_W}.  \label{Zsurvival}
\end{equation}
Since the binding affinity of a PM-specific target to a MM probe is less than to 
a PM probe, we expect in general that $\kappa^\MM > \kappa^\PeM$, or equivalently, 
$s^\MM(t_W) < s^\PeM(t_W)$ and hence $y^\MM(\infty) < y^\PeM(\infty)$ as required.  

Ideally one would like to test directly the veracity of the survival function Eq.(\ref{Zsurvival}) using data from a 
range of washing times.  While this is not possible with spike-in data corresponding to a single value of 
$t_W$, we can at least check for qualitative agreement of the above scenario with probe sequence information.  

From Eqs.~(\ref{y0bKcorrectionW}) and (\ref{Zsurvival}) one obtains 
\begin{equation}
\kappa t_W = \log b_{\rm S} - \log \left[y_0(t_W) + b(t_W) - a\right].  \label{washingrate}
\end{equation}
For fixed $t_W$, the left hand side is a measure of the rate at which probe-target duplexes dissociate 
due to washing, and should increase with decreasing binding affinity.  The right hand side depends on the 
fitted isotherm parameters $y_0(t_W)$ and $b(t_W)$, and two unknown parameters: $a$, the physical background, and 
$b_{\rm S}$, the fluorescence intensity above background of a feature fully saturated with PM-specific probe-target 
duplexes.   In order to make comparisons across the fitted spike-in data, we will take the two unknown parameters 
to be constant across all features of the microarray.  While this may seem to be 
a radical assumption for $b_{\rm S}$, 
we argue that, because the target mRNA is fractionated randomly to lengths of 
between 50 and 200 bases, the distribution of the number of U and C bases carrying biotin labels on 
PM-specific  targets will not be strongly influenced by the relatively short 25-base subsequence of the probe.  The total number of bases carrying labels on a saturated feature could therefore be replaced 
by a typical representative value independent of the feature.  

\begin{figure}
\centering
\includegraphics[scale=1]{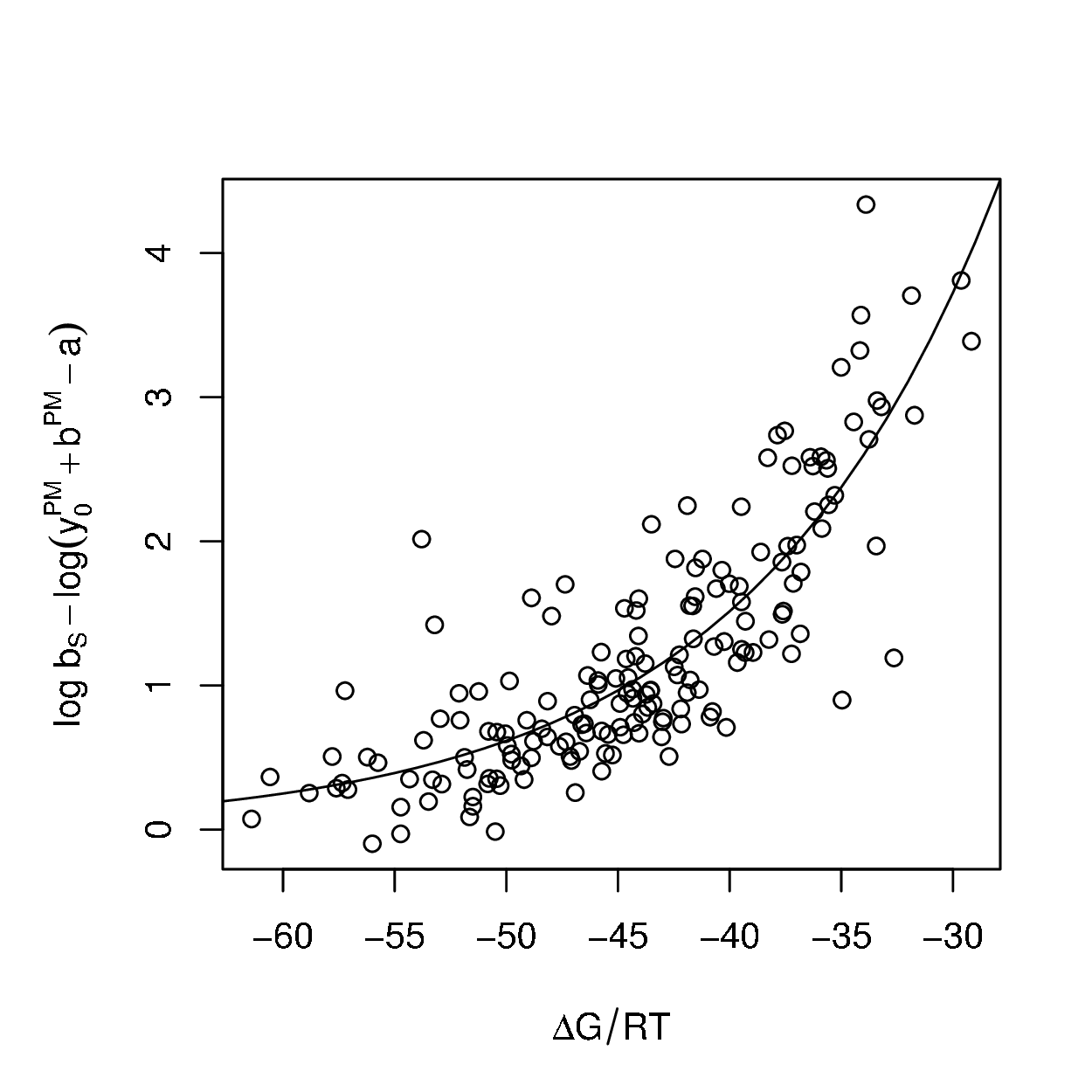}
\caption{Plots of the estimate of the washing rate $\kappa$ (times the washing time $t_W$, which is constant across all proes) given by Eq.~(\ref{washingrate}) using 
Langmuir isotherm parameter fits for the PM probesets of the spike-in experiment described in 
Section~\ref{sec:Langmuir}.  $\Delta G$ is the RNA/DNA duplex free binding energy in bulk 
solution calculated using the nearest neighbour stacking model of ref.~\cite{Sugimoto95}.  
The solid curve is the exponential fit Eq.~(\ref{washingfit}) with parameter values given in the text.}
\label{fig:washing}
\end{figure}

In Fig.~\ref{fig:washing}, we plot the right hand side of Eq.~(\ref{washingrate}) against 
RNA/DNA duplex free binding energy in bulk solution calculated using the nearest neighbour 
stacking model and parameters of ref.~\cite{Sugimoto95}.  Values of $y_0$ and $b$ are 
from fits of the PM data to hyperbolic isotherms as described in Section~\ref{sec:Langmuir}.  The value $a = 50$ was chosen to be slightly less than the lowest intensity value from the entire data set, though in practice any positive value up to 100 gives 
an almost identical plot.  The choice of parameter $b_{\rm S} = 31100$ only affects the vertical offset, and has 
been determined by setting $\log b_{\rm S} = - \alpha$, where $\alpha$ is the intercept in a linear regression to an empirical function 
(also plotted in Fig~\ref{fig:washing})
\begin{equation} 
- \log \left[ y_0(t_W) + b(t_W) - a\right] = \alpha + \beta e^{-\Delta G/(\lambda RT)},  \label{washingfit}
\end{equation}
in which $\lambda = 11.1$ has been chosen to minimize the residual standard error.  The linear regression gives 
$\alpha = 10.3$ and $\beta = 55.4$.  We see that the data is consistent with a rate of duplex removal 
during washing that decreases exponentially to zero with increasing binding energy $-\Delta G$.  
The factor $\lambda$ reflects the fact that effective duplex binding energies at the microarrray 
surface are considerably less than bulk solution binding energies due to effects such as electrostatic 
blocking~\cite{Levicky05} (see also Subsection~\ref{sec:ESBlocking}) and a consequent 
enhancement of partial zippering (see Eq.~(\ref{effectiveK})).  

\begin{figure*}
\centering
\includegraphics[scale=0.6]{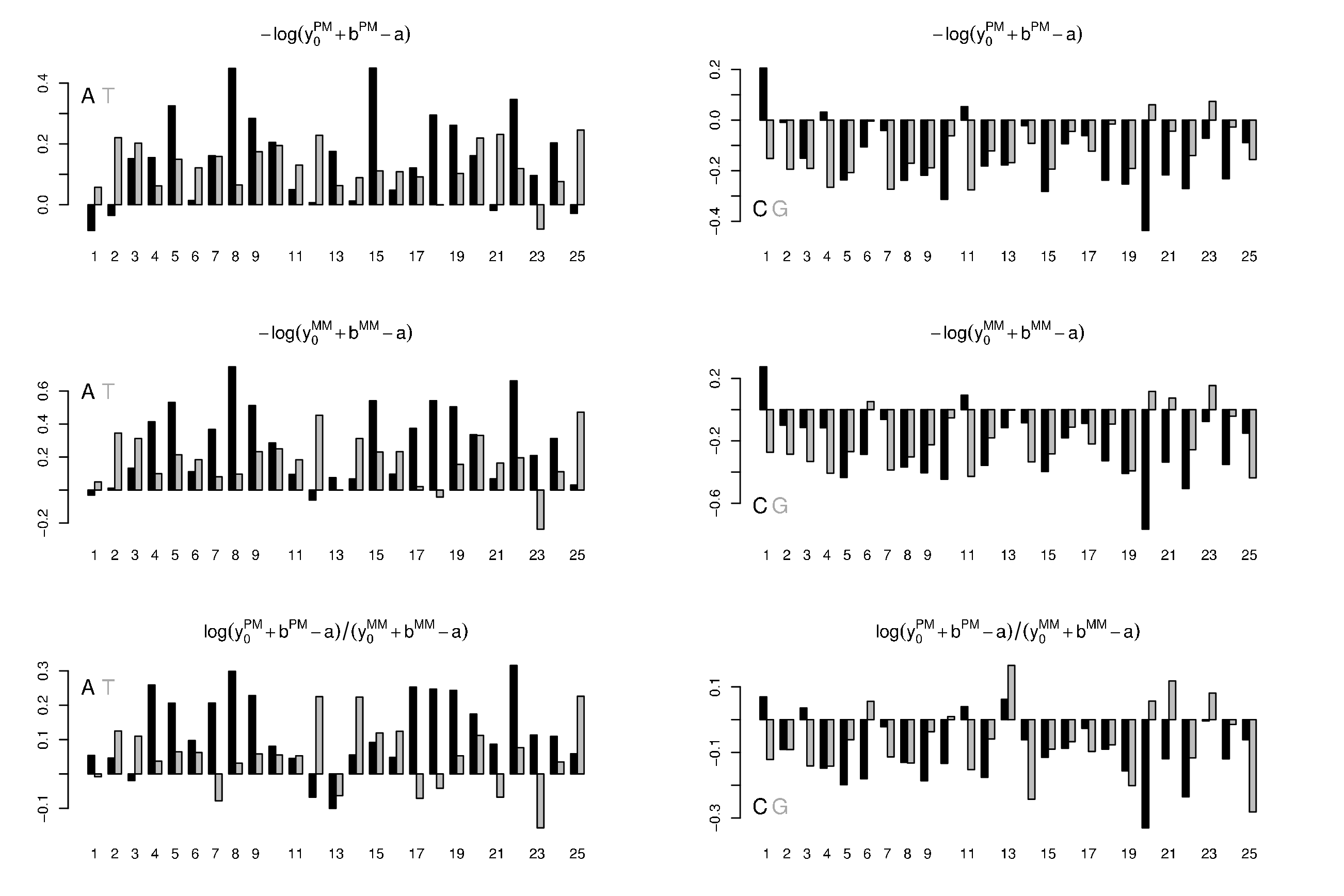}
\caption{Barcharts of estimates of $\kappa^\PeM t_W$ (top), $\kappa^\MM t_W$ (middle) and 
$(\kappa^\MM - \kappa^\PeM)t_W$ (bottom) from Eq.~(\ref{washingrate}) averaged over DNA probe sequences with base A, T, (black, grey respectively, left hand plots), C or G (black, grey respectively, 
right hand plots) at each position along the PM probe sequence minus the corresponding averages 
over all probe sequences. }
\label{fig:barcharts}
\end{figure*}

In Fig.~\ref{fig:barcharts} we examine the dependence of the estimated washing rate Eq.~(\ref{washingrate}) on the 
nucleotide composition of probe sequences.  The upper four bar charts show estimated PM (MM) 
washing rates averaged over sequences with a particular base at the $i$th position ($i = 1,...,25$) 
minus estimated washing rates averaged over all PM (MM) probes.  As expected, washing rates are generally lower than 
average for strong hydrogen bonded bases C and G occurring in the DNA probe sequences and 
higher than average for A and T.  This is the case for both PM and MM probes.  Interestingly, with 
the exception of the mismatched central MM base, there seems to be no obvious relationship 
between the strength of the effect and position along the probe.  
 
 The remaining two bar charts show the analogous contrasts for the difference 
$(\kappa^\MM - \kappa^\PeM) t_W$.  The estimate of this quantity, determined 
from Eq.~(\ref{washingrate}), is independent of $b_{\rm S}$, and so conclusions drawn from from 
this bar chart do not rely on the assumption that $b_{\rm S}$ is uniform from one feature to another.  
Here the effect of the mismatched base at position 13 is quite noticeable: removing a triple 
hydrogen bond (C$\equiv$G) raises the washing rate more than removing a double hydrogen 
bond (A$=$U) or (T$=$A).  Conversely, 
the effect of a central mismatch on the washing rate is almost always greater when any of the remaining 
24 bases is a weakly bound A or T than a strongly bound C or G.  This is entirely in keeping with the 
washing scenario.  

\section{Summary and Conclusions}
\label{sec:Summary}

An understanding of the physical processes driving hybridization is essential if the design 
of expression measures is to advance to a point where target concentration can be measured 
in absolute terms.  The aim of this paper has been to gain an improved understanding of 
the physics of oligonucleotide microarrays by exploiting the observed differences in the 
responses of PM and MM features to known cRNA target concentrations.  
The starting point of this paper is an adsorption model of hybridization 
at the surface of oligonucleotide microarrays based on models proposed independently by 
Hekstra et al.\cite{Hekstra03} and Halperin et al.\cite{Halperin04}.  Though arrived at from different approaches the Hekstra and Halperin models are essentially equivalent, and are an 
improvement on their predecessors in that they allow for the 
presence of cross-hybridization from non-specific targets.  

We have mainly concentrated on seeking to explain the commonly observed difference between 
fluorescence intensity measurements from a neighbouring PM/MM pair of features at high specific 
target concentration.  That is, if a sufficiently high concentration of PM-specific RNA target is 
spiked in to the target solution, both the PM and MM fluorescence intensity signals will reach an 
asymptote, but the MM asymptote is almost invariably observed to be lower than the PM asymptote.  
Our starting Hekstra/Halperin model 
incorrectly predicts 100\% coverage with PM-specific duplexes of both PM and MM 
features under these conditions, which in turn incorrectly implies that the asymptotic PM and MM 
fluorescence signals will be equal.   

We have sought to resolve this discrepancy firstly by taking a more detailed look at the hybridisation 
step, and secondly by examining the subsequent washing step.   In general, we find that more 
detailed variants of our starting model of the hybridisation step, many of which have been 
independently suggested or alluded to previously, are unable to resolve the problem.  
Given our previous analysis of data from the Affymetrix Latin Square spike-in experiment 
\cite{Burden04}, we are able to dismiss the Sips isotherm and non-equilibrium models of 
hybridization including multi-step models which take into account a slow initiation step followed 
by a rapid zipping up.  We are also able to dismiss the effects of electrostatic screening at the 
microarray surface and bulk target-target hybridization as a possible explanation of differential 
PM/MM intensity measurements at saturation.  

We are as yet unable to dismiss entirely the possibility that competitive hybridization from probe-probe 
duplexes at the microarray surface renders a fraction of DNA probes unavailable to target molecules, as 
suggested by Forman et al.\cite{Forman98}.  To make progress with this problem, one needs to carry out a 
numerical simulation of a dimer-like statistical mechanics model on a two dimensional random lattice, 
probably by Monte Carlo methods.  Analysis of the equivalent one dimensional model suggests that this form of 
competitive hybridization could well have a measurable quantitative effect on the equilibrium 
adsorption isotherm, though the two dimensional case is unlikely to lead to the observed hyperbolic 
response curve.  

By comparison, we find that the post-hybridisation washing step is able to provide a promising and straightforward explanation for the PM/MM difference at saturation.  We have considered a scenario in which the equilibrium state predicted by our starting Hekstra/Halperin model is attained by the end of the 
hybridization step, following which the washing phase dissociates a fraction of bound 
duplexes.   The portion of both the PM and MM signals above background decays 
exponentially during the washing phase, 
but since the MM binding affinity is less than that for PM features, the decay rate is faster for MM 
features.   The results of our analysis of the dependence of inferred washing rates on probe base 
sequences support this scenario.  
The advantages of this model are that it preserves the observed hyperbolic shape of the 
Langmuir isotherm, and that it explains both the partial (i.e. $< 100\%$) coverage 
of each feature by duplexes at saturation spike-in concentrations and the fact that the MM feature 
almost invariably asymptotes to a lower measured fluorescence intensity than its PM partner.  

The analysis presented in this paper argues that the solution to providing a practical method of 
estimating absolute concentration of target mRNA from microarray data lies in understanding the 
physics of hybridization and washing at the microarray surface.  Ideally one would like to be able to 
estimate isotherm parameters from probe sequence information and physical parameters including 
microarray design parameters, hybridization temperatures and washing times.  It is 
hoped that theoretical analysis can serve as a guide to the design of experimental work.  In particular, 
the results set out in this paper illustrate a strong need for further spike-in experiments carried out 
with varying washing times or continuous monitoring of fluorescence intensities during the washing step.

\ack
We thank Rodney Baxter and Andrew James for helpful discussions and anonymous referees for helpful input.  This research was partially 
supported by Australian Research Council Discovery Grant DP0343727.

\appendix

\section{Statistical comparison of Langmuir and Sips isotherms}
\label{sec:ASips}

In this appendix we carry out a statistical analysis of fits to the Langmuir isotherm, 
Eq.~(\ref{hypres}), and the Sips isotherm 
\begin{equation}
y = y_0 + b \frac{x^\gamma}{x^\gamma + K^\gamma},       \label{sipsiso}
\end{equation}
to determine which model is the better fit to the MM data of the Affymetrix spike-in experiment.  
The method used is described in detail in an earlier paper which compares fits of the PM data 
to a number of isotherm models\cite{Burden04}.  

The stochastic component of the fluorescence intensity $y$ is assumed to be drawn from a Gamma 
distribution.  The data is fitted using the generalized linear model formalism as defined in 
ref.\cite{McCullagh89}, in which the negative log likelihood of 
the fit, or deviance, is minimized over the parameters $y_0$, $b$, $K$ and, in the case of the 
Sips isotherm, also $\gamma$.  To compare fits to the Langmuir and Sips models with 
$r_L$ and $r_S$ residual degrees of freedom and deviances $D_L$ and $D_S$ respectively, we 
use the scaled deviance  
\begin{equation}
\Delta D_{\rm scaled} = (D_L - D_S) \frac{r_S}{D_S}.   \label{DeltaDscaled}
\end{equation} 
Note that $r_L > r_S >>1$.  To evaluate the null  hypothesis, $\gamma = 1$, $\Delta D_{\rm scaled}$ 
can be compared with a chi-squared distribution with $\Delta r = r_L - r_S$ degrees of 
freedom\cite{McCullagh89}.  

\begin{figure}
\centering
\includegraphics[scale=0.6]{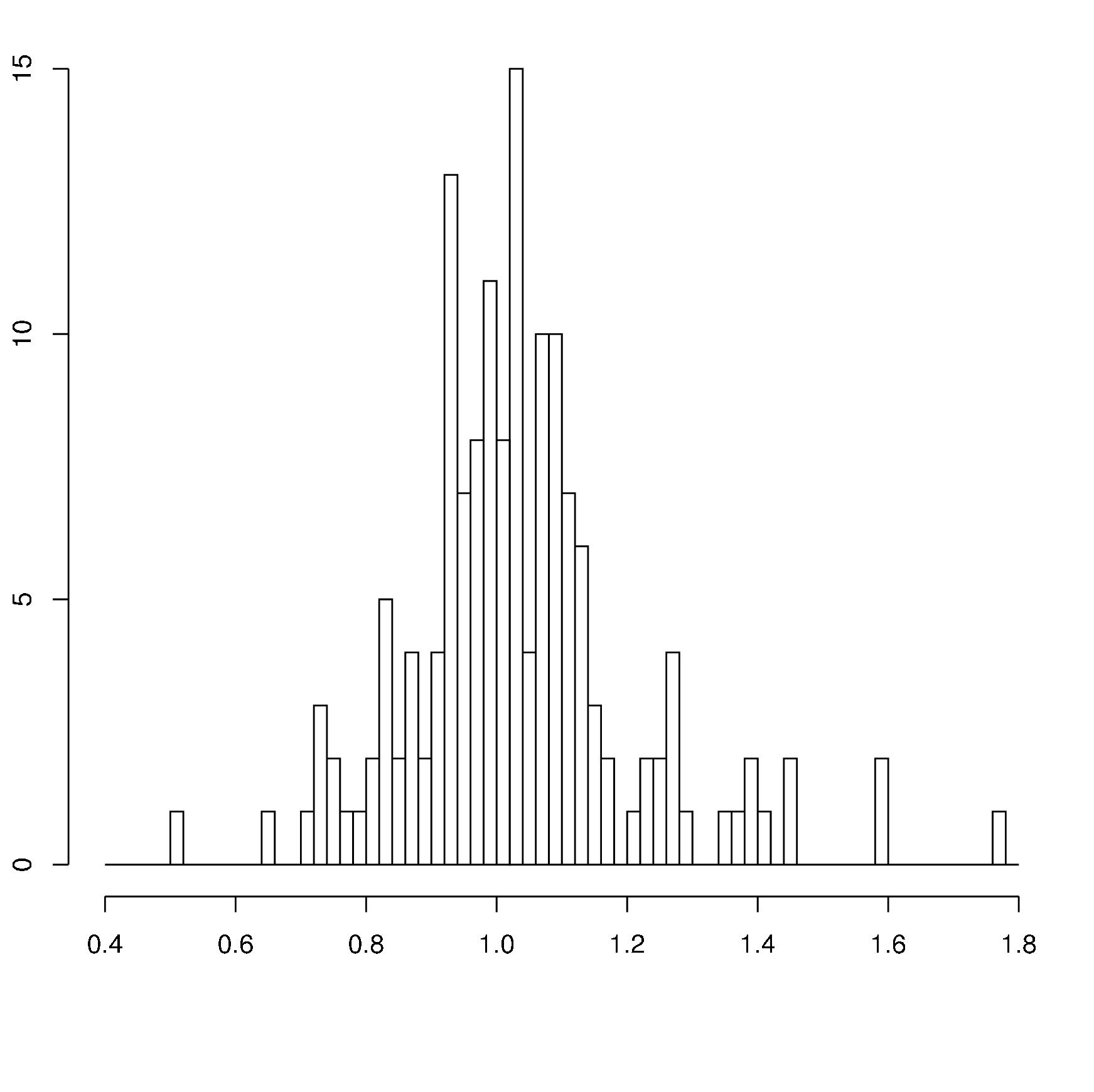}
\caption{Histogram of fitted values of the Sips parameter $\gamma$ for the MM data.}
\label{fig:sipshist}
\end{figure}

We were able to obtain fits with positive parameter values to both the Langmuir and 
Sips isotherms for about 80\% of the probes.  For most of the remaining cases the MM response 
was too small to provide a useful fit.  
Results for the scaled deviance are shown in Table~\ref{tab:Sipscomp}.  The total 
deviance of 133.8 lies at the 13th percentile of a chi-squared distribution with 153 
degrees of freedom, showing no reason to consider a more complex model that the Langmuir isotherm.  
Finally, a histogram of the fitted values of the Sips parameter, Fig.~\ref{fig:sipshist}, shows 
that the Sips parameter is symmetrically distributed about $\gamma = 1$, as expected if the 
Langmuir isotherm is the more accurate model.    

\begin{table}[t]
\caption{\label{tab:Sipscomp} Comparisons of fits to Langmuir and Sips isotherms.  $\Delta r$ is the decrease 
in residual degrees of freedom for each gene and $\Delta D_{\rm scaled}$ is the corresponding 
scaled decrease in deviance from Eq.~(\ref{DeltaDscaled}).  
}

\begin{indented}
\lineup
\item[]\begin{tabular}{lrrl}
\br
   Gene   & $\Delta r$ & $\Delta D_{\rm scaled}$ &   omitted probes \\
\mr
37777\_at &	14  &	 6.43  &	      3, 9      \\	
  684\_at &	12  &	 3.62  &		3, 5, 7, 8   \\
 1597\_at &	12  &	14.56  &		9, 11, 14, 15  \\
38734\_at &	 9  &	10.11  &	      1, 3, 4, 9, 11, 12, 6      \\	
39058\_at &	 5  &	11.34  &		1, 2, 3, 5, 6, 7, 9, 10, 12, 14, 16   \\
36311\_at &	13  &	 3.46  &	      7, 8, 14      \\	
 1024\_at &	16  &	15.19  &	            \\	
36202\_at &	15  &	 6.18  &	      6      \\	
36085\_at &	15  &	 7.29  &	      13      \\	
40322\_at &	16  &	39.58  &	            \\	
 1091\_at &	14  &	 3.83  &	      1, 2      \\	
 1708\_at &	12  &	 2.36  &		11, 12, 13, 14  \\
All genes & 	153  & 133.80  &	            \\	
\br
\end{tabular}
\end{indented}
\end{table}

\section{Quasi-equilibrium model with nucleation}
\label{sec:AQuasi}

\begin{figure}
\centering
\includegraphics[scale=0.6]{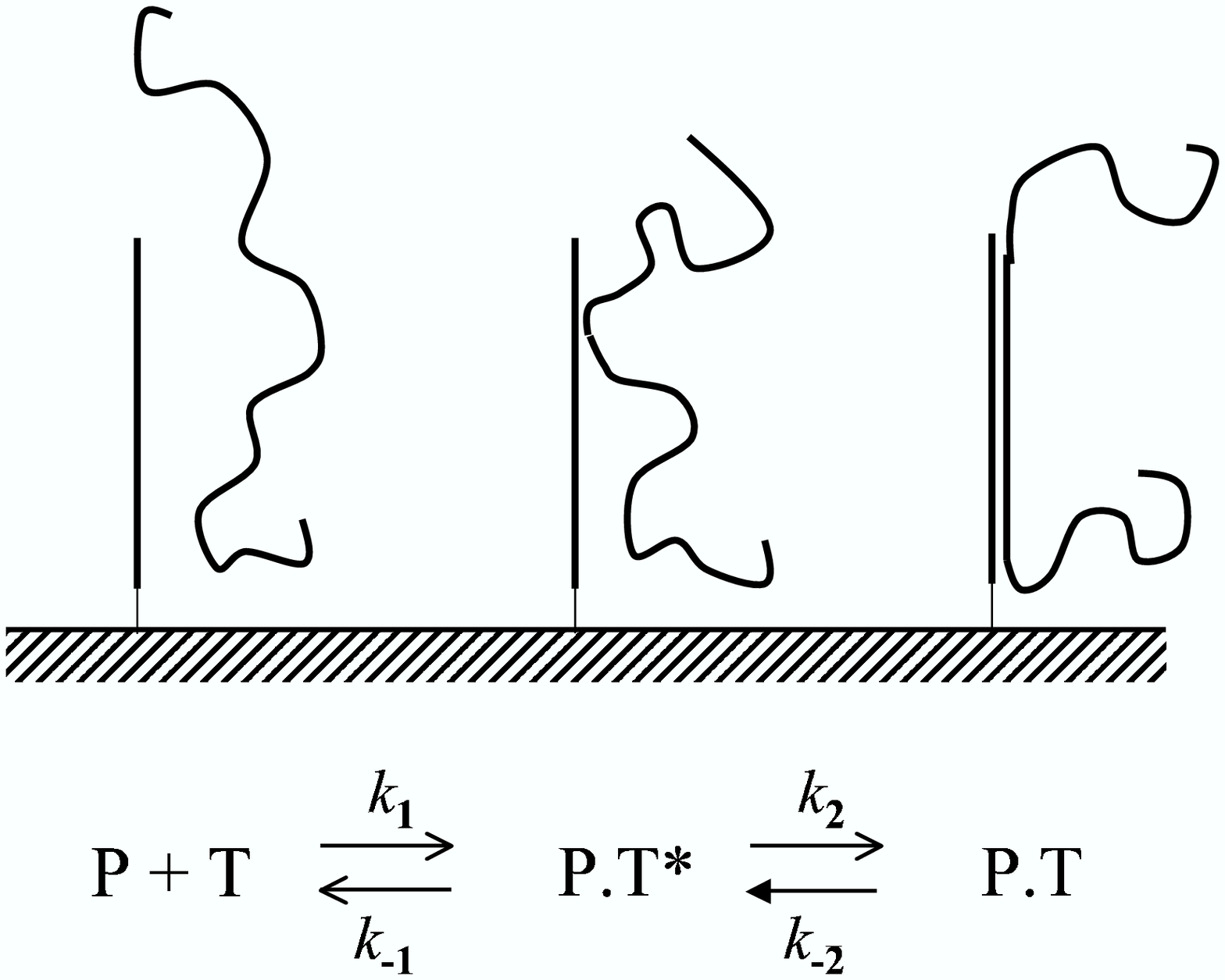}
\caption{Hybridization proceeding from probe plus target (${\rm P} + {\rm T}$) to partially 
formed duplex in which two or three bases pair (${\rm P.T}^*$) to a zipped-up duplex (P.T).  
$k_1$, $k_{-1}$, $k_2$ and $k_{-2}$ are chemical reaction rates. }
\label{fig:zipping}
\end{figure}

We consider the hybridization model illustrated in Fig.~\ref{fig:zipping} in which the forward, 
duplex forming, reaction involves two steps: a slow rate determining step in which the first two or 
three base pairs form, following a fast zipping-up step in which the remaining base pairs form.  
The probe and target molecules are denoted by P and T respectively, the partially formed duplex 
after the rate determining step by P.T$^*$, and the completed target-probe duplex by P.T.  For 
simplicity we consider the case without cross-hybridization.  

Let the target concentration be $x$, the fraction of probes in a feature which have formed a fully 
zipped up duplex P.T be $\theta$ and the fraction which have formed an initiated duplex P.T$^*$ be 
$\zeta$.  The remaining fraction of free single strand probes is defined as $\chi = 1 - \theta - \zeta$.  
The chemical rate equations are
\begin{eqnarray}
\frac{d\chi}{dt} & = & - k_1 x \chi + k_{-1}\zeta, \label{chirate} \\
\frac{d\theta}{dt}   & = & k_2 \zeta - k_{-2} \theta.  \label{thetaratea}
\end{eqnarray}
The reaction rates $k_1 x$ and $k_{-2}$ are assumed to be slow (on the order of hours) and the 
rates $k_{-1}$ and $k_2$ fast.  Accordingly we define 
\begin{equation}
k_1 = \epsilon \kappa_1, \hspace{5 mm} k_{-2} = \epsilon \kappa_{-2}, \hspace{5 mm} 
                 \zeta = \epsilon \hat{\zeta}, 
\end{equation}
where $\epsilon << 1$.  This gives 
\begin{eqnarray}
\frac{d\chi}{d\tau} & = & - \kappa_1 x \chi + k_{-1}\hat{\zeta}, \label{chirate2} \\
\frac{d\theta}{d\tau}   & = & k_2 \hat{\zeta} - \kappa_{-2} \theta,  \label{thetaratea2}
\end{eqnarray}
where $\tau = \epsilon t$ is $O(1)$ on timescales of the slow nucleation reactions.  We solve 
these equations to zeroth order in $\epsilon$, subject to the constraints 
$\theta + \chi = 1 + O(\epsilon)$ and $d\theta/d\tau = - d\chi/d\tau + O(\epsilon)$.  Eliminating 
$\hat{\zeta}$ and $\chi$ with the help of the constraints gives 
\begin{equation}
\frac{d\theta}{d\tau} = \frac{\kappa_1 k_2 x}{k_{-1} + k_2} - 
           \frac{\kappa_1 k_2 x + \kappa_{-2}k_{-1}}{k_{-1} + k_2} \theta + O(\epsilon).  
\end{equation}
The solution to zeroth order, with initial condition $\theta(0) = 0$, is 
\begin{equation}
\theta(t) = \frac{k_1 k_2 x}{k_1 k_2 x + k_{-1} k_{-2}} \left[ 1 - 
    e^{-(k_1 k_2 x + k_{-1}k_{-2})t/(k_{-1} + k_2)} \right], 
\end{equation}
after reinstating the original variables.  This is of the form Eq.~(\ref{tsoln}) where 
$K_{\rm S} = k_{-1}k_{-2}/(k_1 k_2)$ and $k_{\rm f} = k_1 k_2/(k_{-1}+k_2)$.  

\section{Equilibrium model with competition between probe-target and probe-probe duplexes}
\label{sec:APP}

We consider here the equilibrium thermodynamics of the microarray surface when pairwise interactions 
between neighbouring probes and self interaction of individual probes are taken into account.  
The formation of probe-probe duplexes or folded probes will render a fraction of the probes 
unavailable for RNA target hybridization.  

For a given feature, define $M$ to be the total number of probe sites on that feature, $N$ to be the number of probe-target duplexes, $P$ to be the number of probe-probe duplexes and $Q$ to be the 
number of self interacting (i.e. folded) probes.   In this appendix we will for simplicity ignore hybridization of non-specific targets and partial zippering.  The number of 
configurations consistent with the above partitioning is 
\begin{equation}
g(M,N,P,Q) = \frac{\nu(P,M) (M - 2P)!}{N! Q! (M - N - 2P - Q)!},
\end{equation}
where $\nu(P,M)$ is the number of ways of forming $P$ neighbouring pair duplexes on an array of $M$ sites, 
where $0 \le 2P \le M$.  The contribution to the canonical partition function from the entire feature is 
\begin{eqnarray}
\lefteqn{e^{-M\hat{\gamma}/k_{\rm B}T} = g(M,N,P,Q) \times}  & & \label{partitionfunc} \\
 & & \exp \left(\frac{1}{k_{\rm B}T} \left[ \hat{\mu}^0_{\rm pt} N + \hat{\mu}^0_{\rm pp} P 
          + \hat{\mu}^0_{\rm q} Q 
                             + \hat{\mu}^0_{\rm p} (M - N - 2P - Q) \right] \right), \nonumber
\end{eqnarray} 
where $\hat{\gamma}$ is the free energy per site, $\hat{\mu}^0_{\rm pt}$, $\hat{\mu}^0_{\rm pp}$, 
$\hat{\mu}^0_{\rm q}$ and $\hat{\mu}^0_{\rm p}$ are reference state chemical potentials per site 
of a probe-target duplex, probe-probe duplex, self interacting probe and unmatched probe 
respectively, and $k_{\rm B}$ is Boltzmann's constant.  

For illustrative purposes we begin with an analysis of the relatively easily solved one dimensional 
model.  For a one dimensional lattice in which nearest neighbour sites may form duplexes, one easily 
obtains $\nu(P,M) = (M - P)!/\left[ P! (M - 2P)!\right]$, and hence
\begin{equation}
g(M,N,P,Q) = \frac{(M - P)!}{N! P! Q! (M - N - 2P - Q)!}.  
\end{equation}
Appling the Stirling approximation $\log N! = N\ln N - N + O(\ln N)$ and setting 
\begin{eqnarray}
\theta = \frac{N}{M} & = & \mbox{fraction of feature covered by P-T duplexes}  \nonumber \\
\zeta = \frac{2P}{M} & = & \mbox{fraction of feature covered by P-P duplexes}  \nonumber \\
\xi = \frac{Q}{M} & = & \mbox{fraction of feature covered self interacting probes}  \nonumber \\
\gamma = \frac{R}{k_{\rm B}} \hat{\gamma} & = & \mbox{surface free energy per mole of probe sites}
                                                                               \nonumber
\end{eqnarray}
gives, in the bulk limit $M\rightarrow \infty$,
\begin{eqnarray}
\gamma & = & RT \left[ -(1 - \halfzeta) \ln(1 - \halfzeta) + \halfzeta \ln \halfzeta  + \xi \ln\xi + 
                          \theta \ln\theta 
                                                                 \right.   \nonumber \\
                &    &  \left.  +   \left(1 - \theta - \zeta -\xi \right)\ln\left(1 - \theta - \zeta -\xi \right) \right] \nonumber \\
  & & + \; \theta \mu_{\rm pt}^0 + \halfzeta \mu_{\rm pp}^0 + \; \xi \mu_{\rm q}^0 
           + (1 - \theta - \zeta - \xi) \mu_{\rm p}^0,  \label{ppfreeenergy}
\end{eqnarray}
where $\mu^0_{\rm pt}$, $\mu^0_{\rm pp}$, $\mu^0_{\rm q}$ and $\mu^0_{\rm p}$ are reference 
state chemical potentials per mole and $R$ is the gas constant.  

The equilibrium isotherm is obtained by balancing exchange chemical potentials for 
P-T duplexes with the chemical potential of the target species in solution and setting the chemical potentials for P-P duplexes and self interacting probes to zero, that is
\begin{equation}
\frac{\partial \gamma}{\partial \theta} = \mu_{\rm t}, \hspace{5mm}
\frac{\partial \gamma}{\partial \zeta} = 0, \hspace{5mm} \frac{\partial \gamma}{\partial \xi} = 0, 
\end{equation}
where $\mu_{\rm t}$ is given by Eq.~(\ref{targetpot}).  This 
leads to  
\begin{equation}
\theta  =  (1 - \zeta - \xi)\frac{x}{x + K_{\rm S}},    \label{thetaPP}
\end{equation}
\begin{equation}
\halfzeta(1 - \halfzeta) = K_{\rm P} (1 - \theta - \zeta - \xi)^2,   \label{thetazeta}
\end{equation}
and 
\begin{equation}
\xi = \frac{1 - \theta - \zeta}{1 + K_{\rm Q}^{-1}},   \label{xieqn}
\end{equation}
where the equilibrium constants $K_{\rm S}$ for the P-T duplex forming reaction, $K_{\rm P}$ 
for the P-P duplex forming reaction and $K_{\rm Q}$ for the self interaction are 
\begin{equation}
K_{\rm S} = x_0 e^{\Delta G/RT}, \hspace{5 mm}  
K_{\rm P} = e^{- \Delta G_{\rm P}/RT}, \hspace{5 mm} 
K_{\rm Q} = e^{- \Delta G_{\rm Q}/RT}, 
\end{equation}
where 
\begin{equation}
\Delta G = \mu_{\rm pt}^0 - \mu_{\rm p}^0 - \mu_{\rm t}^0, \hspace{5 mm}  
\Delta G_{\rm P} = \mu_{\rm pp}^0 - 2 \mu_{\rm p}^0, \hspace{5 mm} 
\Delta G_{\rm Q} = \mu_{\rm q}^0 - \mu_{\rm p}^0. 
\end{equation}
Eliminating $\xi$ gives
\begin{equation}
\theta  =  (1 - \zeta)\frac{x}{x + K'_{\rm S}},    \label{thetaPPprime}
\end{equation}
\begin{equation}
\halfzeta(1 - \halfzeta) = K'_{\rm P} (1 - \theta - \zeta)^2,   \label{thetazetaprime}
\end{equation}
where $K'_{\rm S} = (1 + K_{\rm Q})K_{\rm S}$ and $K'_{\rm P} = K_{\rm P}/(1 + K_{\rm Q})^2$.  
That is, the effect of probe self interaction is to rescale the remaining equilibrium constants.  

From Eq.~(\ref{thetazetaprime}) one finds that the P-P coverage fraction $\zeta$ decreases 
smoothly from a 
maximum value $\zeta_{\rm max} = 1 - \frac{1}{2}\left(K'_{\rm P} + \frac{1}{4}\right)^{-1/2}$ 
at $\theta= 0$ to zero at $\theta = 1$. This has two consequences.  Firstly, there is no 
phase transition, as expected for a one dimensional model with local interactions.  Secondly, 
we see from Eq.~(\ref{thetaPPprime}) 
that $\theta$ asymptotes to 1 in the limit of high target concentration $x \rightarrow \infty$.  
Thus the simple one dimensional model of P-P duplexes is unable to explain partial saturation of the 
feature at high concentration.  A plot of $\theta$ against target concentration for a range of values 
of $K'_{\rm P}$ is given in Fig.~\ref{fig:onedim}.  

Ideally we need to solve the model defined by Eq.~(\ref{partitionfunc}) for a random two dimensional 
lattice.  The presence of self interactions of individual probes involves no interaction between sites and 
consequently cannot complicate the phase structure.  In fact, by comparing Eqs.~(\ref{freeenergy}) 
and (\ref{ppfreeenergy}) we see that, for the purposes of determining phase structure, probe self interactions and non-specific hybridization are mathematically identical problems.  
Nearest neighbour probe-probe interactions, 
on the other hand, are less tractible.  In this case one needs to calculate $\nu(P,M)$ for a 
random two dimensional lattice with some reasonable definition of `neighbouring'.  To analyze 
the bulk limit, it can be shown that one only needs $(1/M)\log \nu(P,M)$ in the limit 
$M$, $P \rightarrow \infty$ for given fixed $2P/M$.  This is the random lattice analogue of the 
monomer-dimer model which is usually defined on a regular two 
dimensional lattice, and for which no exact solution has been found.  For a square lattice, though,  
numerical calculations strongly suggest the model has no phase transition at non-zero monomer 
density\cite{Baxter68}.  (At zero monomer density, that is $2P = M$, the square lattice monomer-dimer 
model is critical, corresponding to the critical point of the Ising model\cite{Kasteleyn63}.) 

A review of most of the two dimensional statistical models which have been solved exactly 
can be found in ref.~\cite{Baxter82}.  These include the close packed dimer model on a square lattice, which 
is equivalent to calculating $\nu(\frac{1}{2}M,M)$, and the hard hexagon model, in which sites of a 
triangular lattice are occupied subject to the constraint that no two neighbouring sites may be occupied
simultaneously.  The model we are interested in is similar in some ways to the hard hexagon model, 
except that in our case links 
of a lattice are occupied subject to the constraint that no two adjoining links may be simultaneously 
occupied.  The hard hexagon model does undergo a phase transition between a liquid phase (uncorrelated 
positioning of hexagons at low density) and a solid phase (close packing of hexagons centred on one of 
three possible sublattices).  Whether the random lattice duplex model relevant to the case in hand 
undergoes a phase transition from a disordered phase at low duplex density or high temperature to an 
ordered phase at high duplex density or low temperature is unknown.

\end{document}